\documentclass{PoS}

\title{The nova V1369 Cen -- a short review}

\ShortTitle{V1369 Cen}

\author{\speaker{Luca Izzo}\\
        Sapienza University of Rome and ICRANet, Rome, Italy\\
        E-mail: \email{luca.izzo@gmail.com}}

\author{Massimo Della Valle\\
        INAF -- Osservatorio Astronomico di Capodimonte, Napoli, Italy\\
        E-mail: \email{dellavalle@na.astro.it}}

\author{Francesca Matteucci\\
        Dipartimento di Fisica, Universit\'a di Trieste, Trieste, Italy\\}
        
\author{Donatella Romano\\
        INAF -- Osservatorio Astronomico di Bologna, Bologna, Italy\\}
        
\author{Luca Pasquini\\
        ESO, Garching bei Munchen, Germany\\}
      
\author{Leonardo Vanzi, Andres Jordan, Jose Miguel Fernandez, Paz Bluhm, Rafael Brahm, Nestor Espinoza\\
        Institute of Astrophysics and Center of Astro Engineering, PUC-Chile, Santiago, Chile\\}
        
\author{Robert Williams\\
        Space Telescope Science Institute, Baltimore, USA\\     } 

\abstract{We briefly present the spectroscopic evolution of the recent outburst of the classical nova V1369 Cen, and the presence of a narrow absorption line identified as due to the resonance of neutral lithium at 6708 \AA. We also discuss the  consequences for the chemical evolution of lithium in the Galaxy.}

\FullConference{The Golden Age of Cataclysmic Variables and Related Objects - III, Golden2015 \\
		7-12 September 2015\\
		Palermo, Italy}

\begin{document}

On December 2$^{nd}$ 2013 a very bright transient was discovered in the southern sky, in the direction of the Centaurus constellation, reaching in few hours the apparent magnitude of $V \approx 4$. The transient was soon confirmed to be a nova \cite{ATEL5621}, and the progenitor star coincides almost with a $\sim$ 15 magnitude star, which was detected by XMM as an X-ray source \cite{ATEL5628}.

\section{Observations}

\begin{figure}
\centering
\includegraphics[width=.7\textwidth]{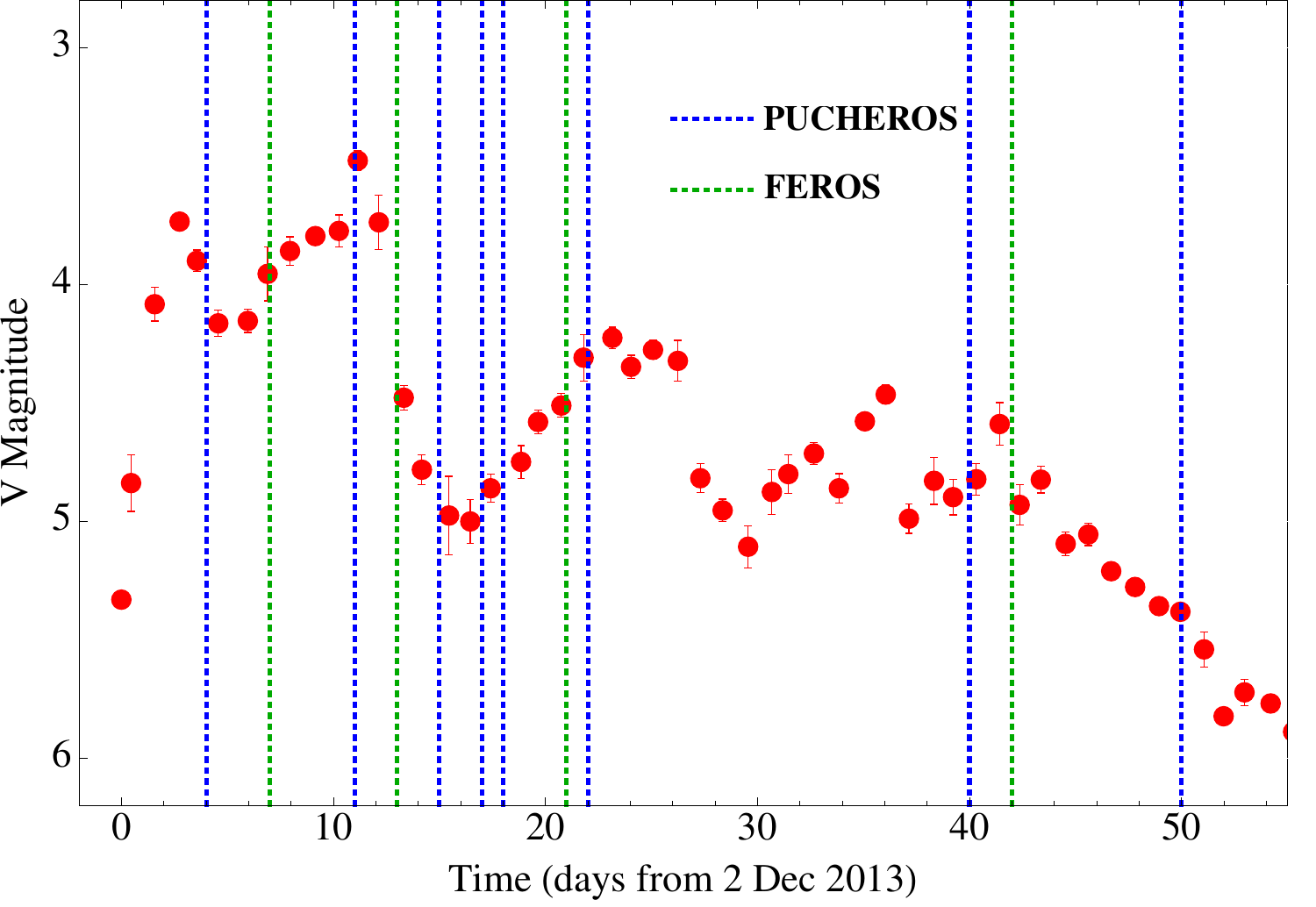}
\caption{The light curve of V1369 Cen in the first 50 days from the initial discovery. $V$-mag data are from the AAVSO archive and have been binned at 1-day cadence. Spectroscopic observations are reported in dashed green (FEROS) and blue (PUCHEROS) lines.}
\label{fig:no1}
\end{figure}

We have started two distinct spectroscopic monitoring programs for the follow-up of this very bright nova\footnote{V1369 Cen is currently (December 2015) the brightest nova of this century.} at the ESO-MPG 2.2 m telescope located on La Silla (Chile) using the Fiber-fed Extended Range Optical Spectrograph (FEROS; R $\sim$ 48000) \cite{Kaufer1999}, and at the Observatory of the Pontificia Universidad Catolica de Chile in Santiago, which consists in a 0.50 m telescope equipped with the PUC High Echelle Resolution Optical Spectrograph (PUCHEROS; R $\sim$ 20000) \cite {Vanzi2012}. The details of the spectroscopic observation in the first 55 days of the nova evolution are shown in Table \ref{tab:no1}, while a light curve, obtained through a rebin of public AAVSO data, and showing the spectroscopic observation epochs is shown in Fig. \ref{fig:no1}.

The early emission of V 1369 Cen is characterized by a rapid increase in luminosity, with a possible total variation from the quiescence of $\Delta_R \approx 12$ \cite{ATEL5621}. After the initial peak of brightness, the light curve presents several rebrightenings, the first of them one week after the initial outburst, with an average luminosity variation of $\Delta V \approx $ 1 mag. Our first spectroscopic observation of V 1369 Cen was carried out 4 days after the initial outburst with PUCHEROS \cite{Izzo2013} and it almost coincides with the first peak of the emission. In the first two weeks the spectra are characterized by the presence of typical Fe II emission lines, suggesting that the nova was already engulfed in its ''iron curtain'' phase \cite{ShoreBASI}. Thanks to the high-resolution of our dataset, we report also the presence of many narrow absorptions, which disappear almost rapidly with time. Details of these narrow absorptions will be presented in the next section. 

\begin{figure}
\centering
\includegraphics[width=.8\textwidth]{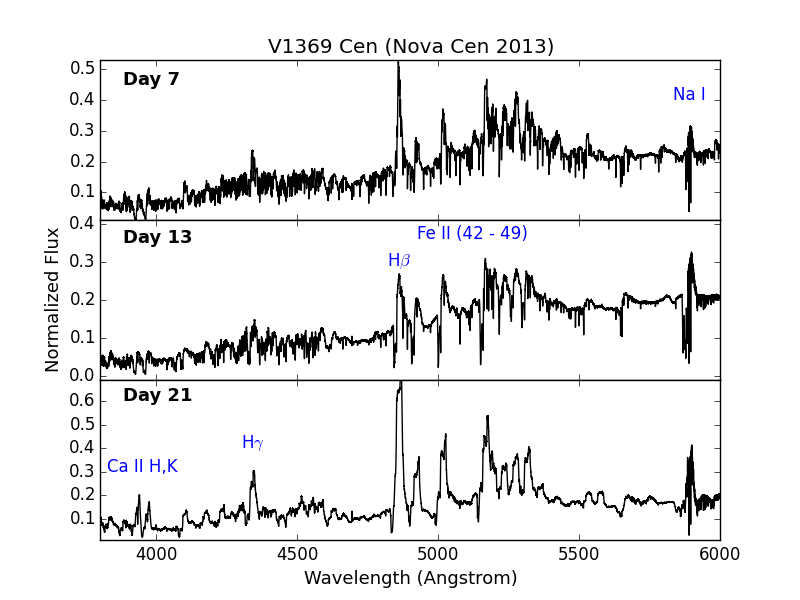}
\caption{The spectral evolution (3800 -- 6000 \AA) of V1369 Cen in the first 21 days from the discovery of the nova.}
\label{fig:no2}
\end{figure}

The optical spectra in the first 14 days of the nova outburst are characterized by strong Balmer emission lines with the Fe II 5169 \AA~the most bright non-Balmer transition, see fig. \ref{fig:no2}. The presence of many absorptions in the range 3700-4600 \AA~complicates the identification of the most common transitions detected in the optically thick phase of novae. We note the presence of distinct expanding velocities for distinct transitions :  for the H$\alpha$ we measure, on Day 4, two expanding systems with average velocities of $\sim$ -500 and -1400 km/s. The same systems are detected for O I 7773-7, 8446 \AA, and for the lower ionization transitions of Fe II (multiplet 42). For the low excitation lines Na I we see initially only the systems expanding at the lower velocity. 

\begin{figure}
\centering
\includegraphics[width=.8\textwidth]{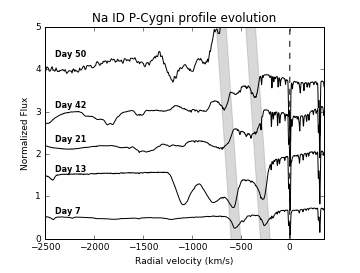}
\caption{The evolution of the P-Cygni profiles of Na I 5890 \AA~during the first 50 days of the nova outburst. The principal absorption is underlined in gray for both the components of the sodium doublet. From Day 11 it is clearly evident the presence of additional absorptions for both components.}
\label{fig:no3}
\end{figure}

A second faster component, observed during and after the first rebrightening (from Day 11) of the nova light curve, is visible in the P-Cygni profiles of all main transitions, including the doublet Na I 5890, see Fig. \ref{fig:no3}. In particular, we note the presence of additional absorptions (what are usually called ''diffuse enhanced absorptions'' \cite{McLaughlin1964}) after each rebrightening suggesting that we are maybe witnessing the presence of multiple ejecta phenomena. Equally important is the evidence that in all epochs it is detected a single principal absorption system, which slightly evolves from an initial velocity of $v_{exp} \sim -500$ km/s reported on Day 7 up to the values of $\sim -800$ km/s, see also Fig. \ref{fig:no3}, which is consistent with the Doppler broadening observed in late nebular emission lines (work in preparation).

\begin{table*}
\centering
\begin{tabular}{lcccccc}
\hline\hline
Observational & Days from & UT start & Exp. time & $\lambda$ range & Resolution & ADC \\
Date & outburst & (hh:mm) & (s) & (\AA\,) & & \\
\hline
\multicolumn{6}{c}{FEROS observations}\\
\hline
09/12/2013 & 7 & 08:10 & 3 $\times$ 40 & 3700-9200 & 48000 & yes \\
15/12/2013 & 13 & 08:31 & 3 $\times$ 25 & 3700-9200 & 48000 & yes \\
23/12/2013 & 21 & 07:34 & 3 $\times$ 45 & 3700-9200 & 48000 & yes \\
13/01/2014 & 42 & 08:38 & 3 $\times$ 60 & 3700-9200 & 48000 & no \\
\hline
\multicolumn{6}{c}{PUCHEROS observations}\\
\hline
06/12/2013 & 4 & 07:36 & 5 $\times$ 300 + 3 $\times$ 600 & 4300-7200 & 20000 & n/a \\
13/12/2013 & 11 & 06:17 & 4 $\times$ 300 + 7 $\times$ 900 & 4300-6800 & 20000 & n/a \\
17/12/2013 & 15 & 07:09 & 3 $\times$ 600 + 1 $\times$ 900 & 4300-7200 & 20000 & n/a \\
19/12/2013 & 17 & 06:57 & 2 $\times$ 600 + 2 $\times$ 900 & 4300-7200 & 20000 & n/a \\
20/12/2013 & 18 & 06:57 & 9 $\times$ 600  & 4300-7200 & 20000 & n/a \\
24/12/2013 & 22 & 07:10 & 4 $\times$ 900 & 4300-7200 & 20000 & n/a \\
11/01/2014 & 40 & 07:33 & 3 $\times$ 600 + 1 $\times$ 900 & 4300-7200 & 20000 & n/a \\
21/01/2014 & 50 & 06:36 & 3 $\times$ 600 + 1 $\times$ 900 & 4300-7200 & 20000 & n/a \\
\hline
\end{tabular}
\caption{Journal of observations of V1369 Cen. The columns report respectively: a) the date and b) the epoch after the nova outburst in days; c) the starting time of the spectroscopic observation in UT; d) the number and duration of the exposure; e) the wavelength range covered; f) the spectroscopic resolution and finally g) if the observations take into account correction for the atmospheric dispersion, i.e., if the ADC mounted on FEROS was operating. }
\label{tab:no1}
\end{table*}

\section{Identification of narrow absorption features and neutral lithium}

\begin{figure}
\centering
\includegraphics[width=.7\textwidth]{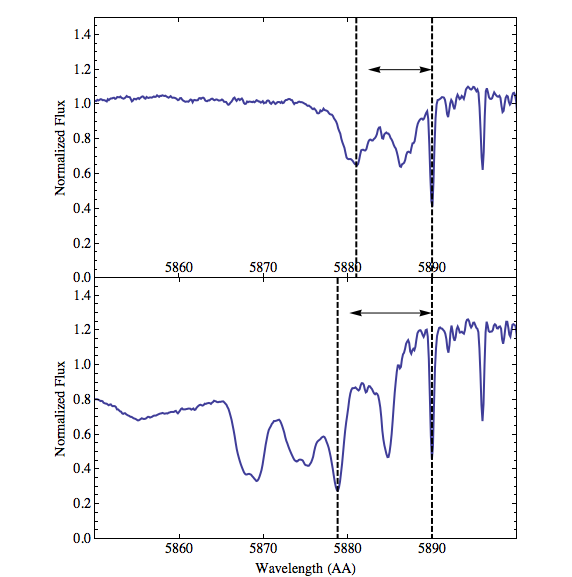}
\caption{The estimate of the expansion velocity for the slower component observed on Day 4 (upper panel) and Day 11 (lower panel) in the P-Cygni profile of the Na I 5890 \AA~line. The $\Delta \lambda$ shift is proportional to the radial expanding velocity through the inverse of the speed of light.}
\label{fig:no4}
\end{figure}

In the first 20 days of observations, we note the presence of hundreds narrow absorption lines, with average FWHM $\approx$ 30-100 km/s, in our spectra. Indeed, in these early epochs the spectra show the nova to be in its ''iron curtain'' phase, clearly marked by to the presence of intense Fe \texttt{II} lines. This means that the ejecta started to cool and start to recombine, while the main responsible for the opacity in the optical range are metal lines \cite{ShoreBASI}. In these epochs it is possible to identify also cool and low-ionization neutral transitions, as the Na \texttt{I} doublet, which is the most clear example. The determination of the principal absorption in P-Cygni profiles of Na I 5890 \AA~allows the determination of the velocity of the expanding gas, where also the metal lines originating narrow absorptions are located. The difference between the P-Cygni slower expanding component and the interstellar lines at rest determines the expanding velocity of the gas, see Fig. \ref{fig:no4} which refers to the case of Na I 5890 \AA~line, while their ratio can be used as multiplicative term for computing expected absorption lines due to other heavier transitions. Following the identification of the narrow absorption lines in Nova LMC 2005, also called transient heavy element absorptions, or THEA \cite{Williams2008}, we have first built a database of neutral and ionized transitions of low-excitation elements and, then applied in each spectrum obtained in the first two weeks of observations, the corresponding factor determined form Na \texttt{I} lines. In this way we have determined about 300 ionized narrow absorption lines and neutral transitions. All these transitions that we have identified are characterized by a low excitation potential, and the kinetic temperature determined from the most abundant elements  (ionized Fe and Ti lines), using the same analysis developed for Nova LMC 2005, is of the order of 10000 K.

\begin{figure}
\centering
\includegraphics[width=.7\textwidth]{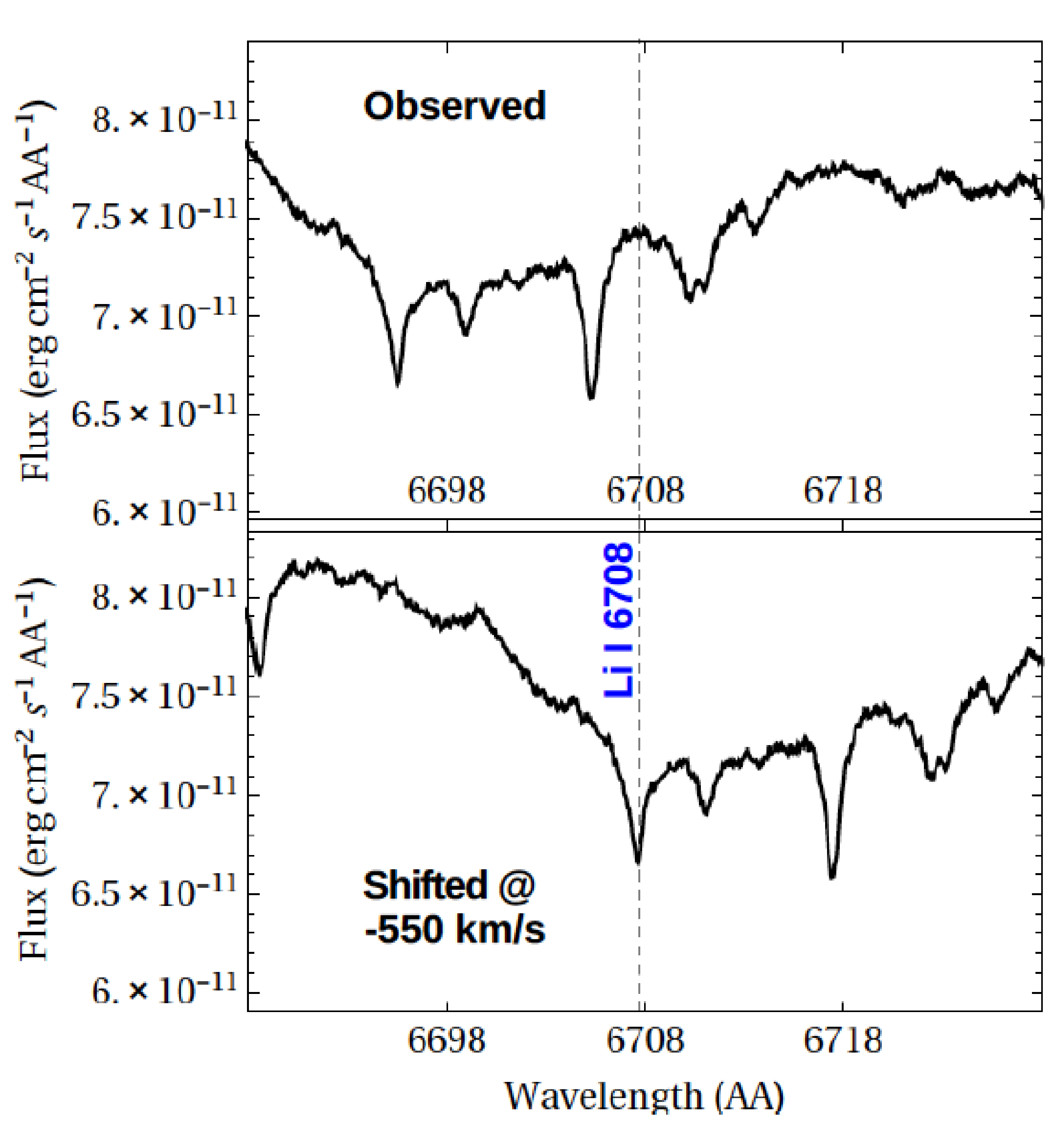}
\caption{The observed spectrum of Day 7 in the wavelength region centered at 6708 \AA (upper panel) and corrected for the velocity of $v_{exp} = -550$ km/s obtained from P-Cygni of Na I lines (lower panel). The identification of the observed absorption at 6695.6 \AA~ as Li I 6707.8 \AA~ is striking. }
\label{fig:no5}
\end{figure}

Among these narrow transitions, we have also observed the presence of a line observed on Day 7 at $\lambda_{obs} = 6695.6 \AA$, which corresponds in the rest-frame to Li \texttt{I} 6708, see Fig. \ref{fig:no5}. Before confirming the identification, we have first tested the association of this line with blue-shifted resonance lithium by verifying that \cite{Izzo2015}: 1) no diffuse interstellar bands are responsible for the observed absorption, and 2) this line is not due to the presence of an over-populated state of common elements, like Fe II or Sc I, due to the UV pumping mechanism \cite{Draine2011}. Therefore we conclude that the most plausible identification for this transition is Li I 6708 \AA. 

\section{Discussions}

The detection of lithium in the spectrum of a classical nova has a twofold interest: first of all, it confirms the Thermo-Nuclear Runaway as the mechanism generating a nova outburst ( see Mariko Kato talk and \cite{Truran1981,Starrfield1978}); secondly, the lithium identification 
confirms novae to be one of the main farms of lithium in the Galaxy \cite{Starrfield1978,Matteucci1991,Romano2001}.

Following Friedjung \cite{Friedjung1979}, since lithium, sodium and potassium are all alkali metals with similar Grotrian diagrams, and ionization energy of resonance transitions, they have formed under similar physical conditions. This assumption means that the relative abundances among these elements can be roughly estimated through the ratio of respective absorption lines  equivalent widths (EWs)\footnote{we must also consider the respective $log\, (gf)$ ratios times a numerical factor dependent on the ratio of the atomic mass units for the elements considered}. However, it is important to remark that this estimate is valid if the energy absorption rate of the atoms considered varies linearly with the number of atoms, i.e. we are in the linear regime of the curve of growth of the elements considered \cite{Spitzer1998}. This assumption seems viable for lithium and potassium, where no saturation is observed in the corresponding blue--shifted absorption lines, while for sodium a deep analysis is needed to confirm the validity of our assumption. However the similarity of the profile shapes of sodium lines with the potassium ones, and partly also with the lithium, seems to exclude saturation effects for sodium lines. 

We have improved our measurements of the EWs of lithium, sodium and potassium, observed in the spectra of Day 7 and Day 11, see Table \ref{tab:no2}. We then use the tabulated values of nova mass ejecta for different CO white dwarf composition \cite{JoseHernanz1998}, in association with the abundance ratios, to compute the mass of lithium ejected in terms of the hydrogen mass. By measuring the intensity of Balmer lines in the nebular phase is possible to estimate the mass of hydrogen ejected \cite{Mason2005}, which results to be in this case of the order of $M_{H,ej} = 10^{-4} M_{\odot}$. We finally obtain a value for the total mass of ejected lithium in V1369 Cen of $M_{Li} =  (0.2, 4.1) \times 10^{-10} M_{\odot}$. After assuming a typical slow nova rate of 15-24 events/yr, this correspond to a total lithium mass ejected in the Galaxy of $M_{Li,ej} = 0.7 - 15 M_{\odot}$, which is in agreement with theoretical estimates \cite{JoseHernanz1998}, and then confirms that novae are one of the main contributors of lithium in the Galaxy.

\begin{table*}
\centering
\begin{tabular}{lccc}
\hline\hline
Epoch & EW(Li 6708) & EW(Na ID) & EW(K I)  \\
(Day) & (\AA\,)  & (\AA\,)  & (\AA\,)  \\
\hline
7 & 0.10 & 2.07 & 0.56\\
11 & 0.08 & 2.33 & 0.68\\
\hline
\end{tabular}
\caption{Equivalent widths measured in the Day 7 and Day 11 spectra for the line Li I 6708 \AA, the Na I doublet 5890, 5896 \AA~and the K I doublet 7665, 7699 \AA.}
\label{tab:no2}
\end{table*}


\begin{thebibliography}{99}
\bibitem{ATEL5621} 
K. Hornoch,
\emph{ATEL} {\bf 5621} (2013)

\bibitem{ATEL5628}
E. Kuulkers, J. U. Ness, A. Ibarra, et al.,
\emph{ATEL} {\bf 5628} (2013)

\bibitem{Kaufer1999}
A. Kaufer, O.Stahl, S. Tubbesing, et al.,
\emph{Msngr} {\bf 95} 8 (1999)

\bibitem{Vanzi2012}
L. Vanzi, J. Chacon, K. G. Helminiak, et al.,
\emph{MNRAS} {\bf 424} 2770 (2012)

\bibitem{Izzo2013}
L. Izzo, E. Mason, L. Vanzi, et al. 
\emph{ATEL} {\bf 5639} (2013)

\bibitem{ShoreBASI}
S. N. Shore, et al.,
\emph{BASI} {\bf 40} 185 (2012)

\bibitem{McLaughlin1964}
D. B. McLaughlin,
\emph{AnAp} {\bf 27} 450 (1964)

\bibitem{Williams2008}
R. Williams, E. Mason, M. Della Valle, et al.,
\emph{ApJ} {\bf 685} 451 (2008)

\bibitem{Izzo2015}
L. Izzo, M. Della Valle, E. Mason, et al.,
\emph{ApJL} {\bf 808} 1 (2015)

\bibitem{Draine2011}
B. T. Draine,
\emph{Physics of the Interstellar and Intergalactic Medium}
Princeton University Press (2011)

\bibitem{Truran1981}
J. W. Truran,
\emph{PrPNP} {\bf 6} 177 (1981)

\bibitem{Starrfield1978}
S. Starrfield, J. W. Truran, W. M. Sparks, et al.,
\emph{ApJ} {\bf 222} 600 (1978)

\bibitem{Matteucci1991}
F. D'Antona, F. Matteucci, 
{A\&A} {\bf 248} 61 (1991)

\bibitem{Romano2001}
D. Romano, F. Matteucci, P. Ventura, et al.,
\emph{A\&A} {\bf 374} 646 (2001)

\bibitem{Friedjung1979}
M. Friedjung,
\emph{A\&A} {\bf 77} 357 (1979)

\bibitem{Spitzer1998}
L. J. Spitzer,
\emph{Physical processes in the Interstellar Medium}
Weinheim, Wiley (1998)

\bibitem{JoseHernanz1998}
J. Jose, M. Hernanz,
\emph{ApJ} {\bf 494} 680 (1998)

\bibitem{Mason2005}
E. Mason, M Della Valle, R. Gilmozzi,
\emph{A\&A} {\bf 435} 1031 (2005)

\end{thebibliography}
\end{document}